# M32: Is There An Ancient, Metal-Poor Population?

G. Fiorentino, A. Monachesi, S. Trager, T. Lauer, A. Saha, K. Mighell, W. L. Freedman, A. Dressler, C. J. Grillmair & E. Tolstoy

*Kapteyn Institute, Postbus 800,9700 AV,Groningen, The Netherlands*
*NOAO, OCIW, Spitzer Science Center*

**Abstract.** We observed two fields near M32 with the ACS/HRC on board the Hubble Space Telescope, located at distances of about 1.8' and 5.4' (hereafter F1 and F2, respectively) from the center of M32. To obtain a very detailed and deep color-magnitude diagram (CMD) and to look for short period variability, we obtained time-series imaging of each field in 32-orbit-long exposures using the F435W (B) and F555W (V) filters, spanning a temporal range of 2 days per filter. We focus on our detection of variability on RR Lyrae variable stars, which represents the only way to obtain information about the presence of a very old population (larger than 10 Gyr) in M32 from optical data. Here we present results obtained from the detection of 31 RR Lyrae in these fields: 17 in F1 and 14 in F2.

**Keywords:** Local Group- M32, stars: RR Lyrae.

**PACS:** 98.52-97.30

## INTRODUCTION

Messier 32 (M32) is the only elliptical galaxy close enough to possibly allow direct observation of its stars down to the main-sequence turn-off (MSTO). It is a vital laboratory for deciphering the stellar populations of all other elliptical galaxies, which can only be studied by the spectra of their integrated light. Major questions about M32's star formation history remain unanswered. M32 appears to have signature of an intermediate age (3-5 Gyr) and close to solar metallicity ([Fe/H]=-0.25 dex) population and an age and a metallicity gradients (Rose 1985, 2005; Trager et al. 2000), but little is known about M32's ancient population (see, e.g., Brown et al. 2000). This conclusion rests on painstaking and controversial spectral analysis of their integrated light. In principle, the most direct information comes from applying stellar evolution theory to color-magnitude diagrams (CMDs), but these common techniques are not applicable to M32 because of the extreme crowding of this compact object. In fact, even the high spatial resolution of the ACS/HRC fails in the search of the old MSTO (Monachesi et al. 2009 in preparation, hereafter M09). Thus, the analysis and the characterization of old stellar population tracers, as RR Lyrae stars, if any, becomes mandatory. Some studies have been already published on M31 Halo and Globular Clusters RR Lyrae population (see Clementini et al. 2001, Brown et al 2004, hereafter B04), but only few out of them have been focused in the neighborhood of M32.

Alonso-Garcia, Mateo & Worthey (2004, hereafter A04) were the first to attempt to directly detect RR Lyrae in fields near M32. They imaged a field about 3.9' (F5) with WFPC2 to the east of M32 and compared it with a control field (F6) well away from M32 that should sample the M31 field stars. They identified variable stars claimed to be RR Lyraes belonging to M32 and therefore suggested that M32 possesses a population that is older than 10 Gyr. They were however unable to classify the RR Lyraes and could not derive periods and amplitudes for them. We have evaluated the completeness of their sample (M09) and believe that A04 only observed about the 20% of the real RR Lyrae sample. This is also supported by a very recent paper (Sarajedini et al. 2009, hereafter S09) where ACS/WFC parallel imaging of our present data set have been used to look for RR Lyrae stars. The two fields are located at 5.3' (F3) and 9.2' (F4) from the center of M32. They found 755 RR Lyrae variables, with excellent photometric and temporal completeness. These RR Lyrae were roughly equally distributed between the two fields, with the same mean average magnitudes, metallicities, and Oosterhoff types in each field. It was therefore impossible for them to separate the variables into M31 and M32 populations. Thus, *it is still therefore an open question as to the precise nature or even presence of RR Lyrae in M32.*

| TABLE 1. Log of Observations | | | | |
|---|---|---|---|---|
| Field | Ra ; Dec (J2000) | Filter | Exposure Time (sec) | Date |
| F1 | 00 42 47.63   +40 50 27.40 | F435W/F555W | 16x1279 +16x1320 | Sept 20-25, 2005 |
| F2 | 00 43 7.89    +40 54 14.50 | F435W/F555W | 16x1279 +16x1320 | Feb 6-12, 2006 |

## Observation And Data Reduction

We observed two fields located at 1.82' (F1) and 5.4' (F2) from the center of M32 with the ACS/HRC on board HST during Cycle 14 (ProgramGO-10572, PI: T. Lauer), as shown in Fig. 1. Time-series imaging of each field were obtained in 32 different orbits using the F435W (B) and F555W (V) filters. The exposure time of each image is nearly 22 minutes. For each field and filter the time window is spread over 2--3 days as shown in Table 1, appropriate for searching for RR Lyrae variables with periods 0.2-1 d.

### *Photometry*

Our optimal photometry has been obtained by using very deep, super-resolved images obtained by combining all 32 images for each filter and field. We have performed photometry by using both the DAOPHOT II/ALLSTAR packages (Stetson 1987, 1994) and by first deconvolving those combined images with a reliable PSF and then performing aperture photometry on the deconvolved images. Both methods returned comparable results and allow us to present the deepest CMD of M32 obtained so far (M09).

The study of the presence of RR Lyr stars is based on a detailed analysis of the time series of our fields. We analyzed each single epoch image (32 per field and per filter) and not a combination of all the images. Because of the intrinsic brightness of the RR Lyrae (V~25 mag), we decided to perform PSF-fitting photometry over all the fully calibrated data products (FLT) images using the DOLPHOT package, a version of HSTphot (Dolphin 2000) modified for ACS images. Our choice has been justified by the short time consumed to obtain high quality (at the RR Lyrae magnitude level) photometry for our data set.

Following the DOLPHOT User's Guide, we have performed the pre-processing steps *mask* and *calcsky* routines before running DOLPHOT. Finally, this package has permitted us to perform photometry simultaneously over all 64 images of each field, returning a catalogue of more than 20000 stars per field already corrected for charge transport efficiency (CTE), for aperture correction and calibrated by following the suggestions by Sirianni et al. (2005). In what follows we will use this photometry to perform the analysis of variable stars.

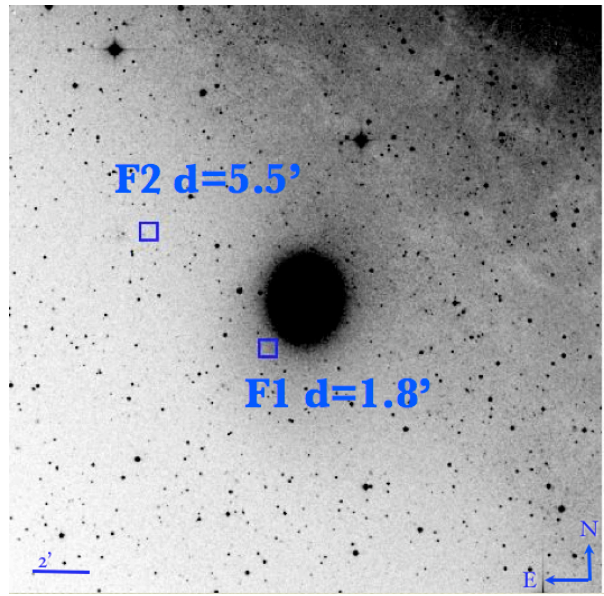

**FIGURE 1.** Finding Chart where fields F1 and F2 have been shown, the ID of the fields and the distance from the center of M32 in arcmin have been labeled in the Figure.

### *Variable Stars Detection And Analysis*

To identify variable sources in both fields, we used a code written by one of us (AS) in the Interactive Data Language (IDL), whose principles are discussed in Saha & Hoessel (1990). This code was applied to the results from DOLPHOT PSF-fitting photometry. Its output gives us not only a list of candidate variable stars but also a good initial estimate of the period. The method assumes that realistic error estimates for each object at each epoch are available from the photometry, which is first used to estimate a chi-square based probability that any given object is a variable. A list of candidates is so chosen, and each candidate is then tested for periodicity and plausible light curves. The graphical interface of this program clearly shows possible aliases, and allows the user to examine the light curves implied for each such alias. The final decision making is done by the user.

A refinement of the period for all the candidate variables has been performed by using two other independent codes. First we used the Period Dispersion Minimization (PDM) alghorithm, in the IRAF environment, to confirm the found periodicity (Stellingwerf 1978). Further refinement was then obtained by using GRATIS (GRaphical Analyzer of TIme Series, a private software developed by P. Montegriffo at the Bologna Observatory; see Clementini et al. 2000 for details), which allows fitting of Fourier series to the magnitudes in each passband as a function of their phase. To calibrate these onto the Johnson-Cousins (JC) system, we need to take into account the color variation in a periodic cycle of variable stars. We thus associated each phased epoch in F435W filter to the corresponding best-fitting F555W model provided by GRATIS at that epoch, and vice versa in an iterative process. Finally, we applied Sirianni et al. (2005) transformations to bring the ACS/HRC magnitudes onto the JC system. This procedure has allowed us to derive well-sampled and consistent light curves in both filters and therefore proper periods, mean magnitudes, colors, and amplitudes, for a total number of 31 *bona fide* RR Lyrae stars: 17 in F1 and 14 in F2. The RR Lyrae locations in the CMDs is shown in Figures 2 and 3, where the averaged magnitudes in HST VEGA mag have been used. Details about the analysis can be found in Fiorentino et al. (2009 in preparation, hereafter F09). As example, we show in Figures 4 and 5 the Atlas of the light curves in V bands obtained for the RR Lyrae belonging to F1 and F2, respectively. The mean V-band magnitude as well as the mean period of this sample have also been labeled.

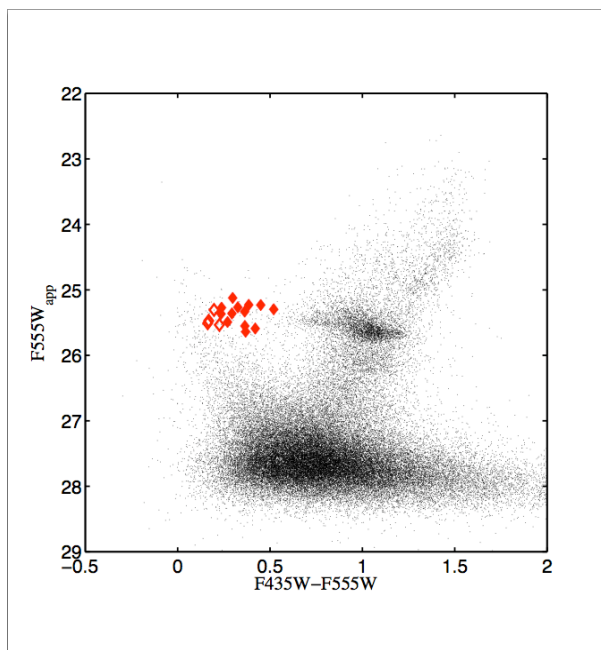

**FIGURE 2.** The CMD of F1 obtained by M09 where the location of the RR Lyrae have been shown. The averaged magnitudes have been properly weighted on the light curve. Empty and filled diamonds represent FO and FU pulsators respectively.

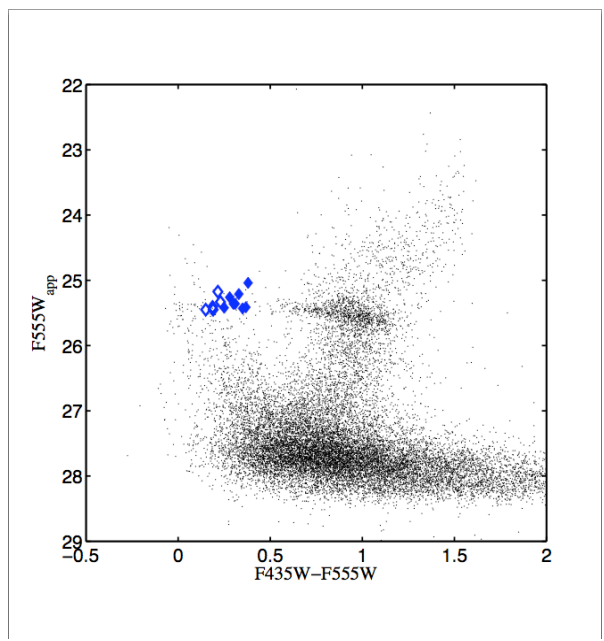

**FIGURE 3.** The same than in Figure 2 but for field F2.

Here we want to stress that this is the first time such detailed light curves have been determined at this crowding level, confirming once again the irreplaceable performance of an instrument such as the HRC in the context of resolved stellar population studies.

## RR Lyrae properties

We now ask whether the detected RR Lyrae stars belong to the same (either the halo of M31 or M32) or different stellar populations (both the halos of M31 and M32) by studying their intrinsic pulsation properties. In fact, as stressed before, the main goal of the present paper is to find reliable evidence of the presence of an ancient stellar population in M32.

### Mean Magnitudes, Periods, Amplitudes and Oosterhoff Types

In according to the analysis described in the previous section, the estimated mean magnitudes are $<V>=25.34 \pm 0.15$ mag and $<V>=25.30 \pm 0.12$ mag for F1 and F2, respectively. Moreover, although our sample of RR Lyrae variables is not large enough to determine statistical properties, we find a mean period for fundamental mode pulsators $<P_{ab}>=0.59\pm0.11$ d for both fields and the two different ratios between FO and FU pulsators: $N_c/N_{ab}=0.36$ and $N_c/N_{ab}=0.62$ for F1 and F2, respectively. The error on the mean periods is just the standard deviations of the average value

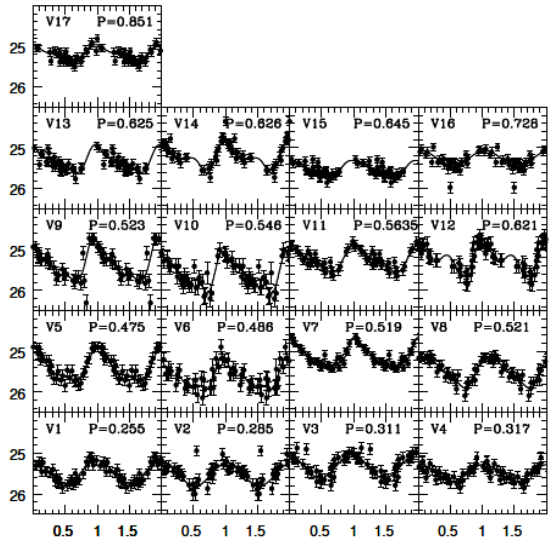

**FIGURE 4.** V band light curves for the 17 RR Lyrae stars found in F1, the ID and the period for each variable is labeled in each panel. The errors bars represent both photometric errors and the residual errors from the model obtained by GRATIS.

computed on our very small sample of RR$_{ab}$ Lyraes, i.e., 14 stars for F1 and 8 for F2.

In Figure 6, we compare the observed properties of the RR Lyrae stars we find in the Bailey diagram with larger sample observed with HST in the surroundings of M32, namely F3 and F4 (S09) and the field from Brown et al. (2004: hereafter F7, see the next section), we find a very good agreement of all the RR Lyrae stars classifiable on this basis as an OoI type.

We estimate the metallicity of the RR Lyrae stars using the relation between luminosity, period and amplitude in V-band found by Alcock et al. (2000) for the LMC, on the Zinn & West (1984) metallicity scale. The uncertainty of this relation is ~ 0.31. By applying this relation to our two samples we find [Fe/H] = -1.52 and -1.65 (± 0.08 ± 0.48 dex) for F1 and F2 respectively. The first errors take into account the uncertainties of both periods and amplitudes, whereas the second one are just the standard deviation from the mean assumed value.

Then we use the Luminosity-Metallicity relation discussed in Cacciari & Clementini (2003) to estimate the distance modulus, assuming E(B-V)=0.08 and [Fe/H]=-1.6 dex. We find a value of 24.53±0.21 mag and of 24.49±0.19 mag for F1 and F2 in very good agreement with estimates already presented in literature for M31 (e. g. Ajhar et al. 1996).

## RR Lyrae nearby M32

By examining all of the pulsation properties of our detected RR Lyrae stars, we conclude that these two groups of stars do not present any significant statistical difference. In this section, we will try to understand whether some of our detections can be associated to M32.

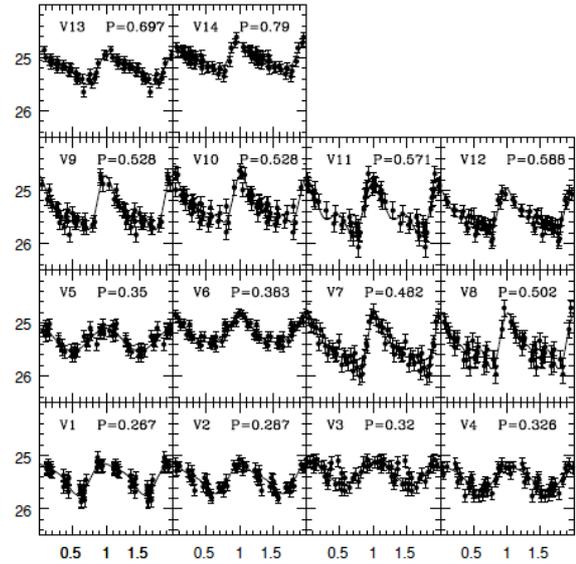

**FIGURE 5.** The same that in Figure 4, but for F2.

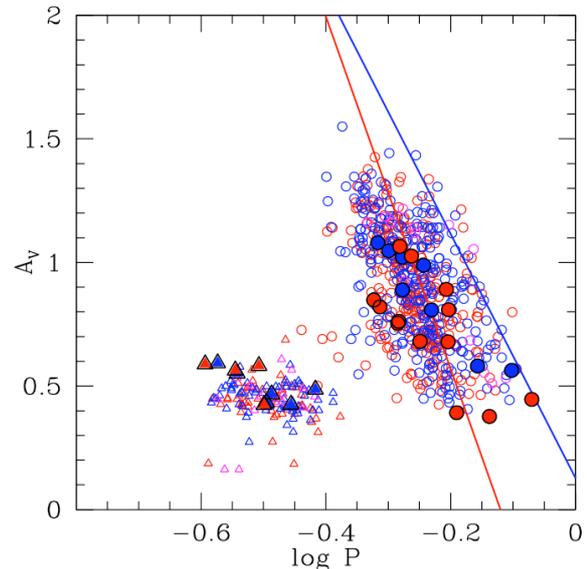

**FIGURE 6.** Comparison in the Bailey plane (amplitude vs period) between the RR Lyrae observed in F1 and F2 (large filled symbols red and blue, respectively) and previous detections nearby M32. The empty symbols represent the larger sample found by S09 (blue and red) and B04 (magenta) by using the ACS/WFC. Triangles and circle represent FO and FU respectively.

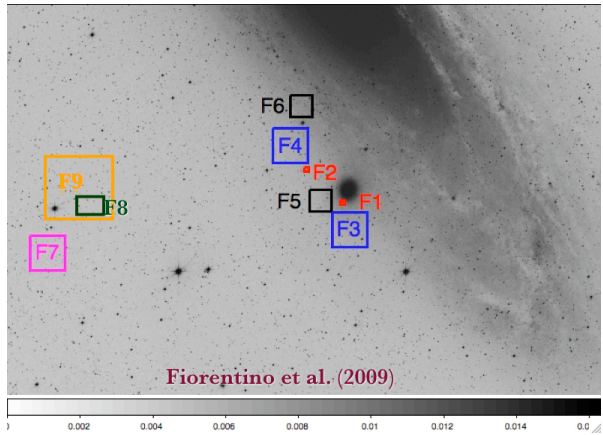

**FIGURE 7.** Fields close to M32 where RR Lyrae have been observed to date.

We begin by collecting all fields in the surroundings of M32 where RR Lyrae stars have been found to date [HST sample: F1-F2 (HRC), from F09; F3-F4 (WFC), from S09; F5-F6 (WFPC2), from A04; F7 (WFC), from B04; Ground Based sample: F8, from Pritchet et al. 1987 and F9, from Dolphin et al. 2004]. The locations of the fields we collect are shown in Figure 7. By assuming a uniform distribution of the RR Lyrae across each field, we compute the $N_{RR}$ per arcmin$^2$:

- 68±18 for F1 and 56±16 F2, assuming 90% completeness;
- 40±7 for F3 and 36±7, assuming 85% completeness;
- 29±40 and 20±36, for F5 and F6. These fields cannot really be considered because of the extremely low completeness estimated to be about 15%, due mainly to two factors: low spatial resolution of the WFPC2 as well as a small time window (about 5 hours);
- 6±1 for F7, assuming 95% completeness;
- 4±12 and 0.2±0.9, respectively for both F8 and F9 (ground based data).

The errors take into account both photometric and temporal completeness as well as Poisson error. This last error is significant only for F1 and F2.

Because of the very small FoV (~ 0.25 arcmin$^2$) of HRC/ACS, the Poisson errors suggest that F1, F2 and F3 have the same number of RR Lyrae stars within 1σ and F4 as well within 2σ. Thus, although we might speculate about the possible slight excess of RR Lyrae at the location of M32, it is impossible to statistically distinguish them from those belonging to M31 Halo.

Unfortunately, despite the high spatial resolution of HRC/ACS, the small size of its FoV does not allow us to claim that we are observing RR Lyrae stars from M32.

Concluding, we think that the region nearby M32 undoubtedly deserves further investigation concerning the RR Lyrae population, to address whether these stars represent the ancient population of M32 or just the halo of M31. Thus the next step should be to overcome this statistical deficiency by using, e. g. ACS/WFC for similar crucial pointings. In fact this camera has already shown (S09) a high performance into detecting RR Lyrae with a good photometric completeness in the presence of an almost prohibitive crowding.

## ACKNOWLEDGMENTS

G. F. wishes to thank M. Irwin, G. Clementini, E. Bernard, M. Monelli and R. Contreras for interesting discussion about this work.